\documentclass[epj]{svjour}

\usepackage{bm}
\usepackage{graphics}
\usepackage{amsfonts}
\usepackage{amssymb}
\usepackage{amsmath}

\begin{document}

\title{Reading entanglement in terms of spin configurations in quantum magnets}

\author{Andrea Fubini \inst{1,2} \and Tommaso Roscilde \inst{3} \and 
Valerio Tognetti\inst{2,4,6} \and Matteo Tusa \inst{2} \and Paola Verrucchi
\inst{4,5}}

\institute{MATIS CNR-INFM \&
Dipartimento di Metodologie Fisiche e Chimiche -
Universit\`a di Catania,
V.le A.~Doria 6, I-95125 Catania, Italy \and 
Dipartimento di Fisica dell'Universit\`a di Firenze,
Via G. Sansone 1, I-50019 Sesto F.no (FI), Italy \and 
Department of Physics and Astronomy, University of Southern
California, Los Angeles, CA 90089-0484 \and 
CNR-INFM, UdR Firenze, Via G. Sansone 1, I-50019 Sesto F.no (FI), Italy \and
Istituto dei Sistemi Complessi - CNR, 
Sez. di Firenze via Madonna del Piano, I-50019 Sesto F.no (FI), Italy \and
Istituto Nazionale di Fisica Nucleare, Sez.
di Firenze, Via G. Sansone 1, I-50019 Sesto F.no (FI), Italy}

\date{Received: date / Revised version: date}

\abstract{
We consider a quantum many-body system made of $N$ interacting
$S{=}1/2$ spins on a lattice, and develop a formalism which allows to
extract, out of conventional magnetic observables, the quantum
probabilities for any selected spin pair to be in maximally entangled
or factorized two-spin states.  This result is used in order to
capture the meaning of entanglement properties in terms of magnetic
behavior.  In particular, we consider the concurrence between two
spins and show how its expression extracts information on the presence
of bipartite entanglement out of the probability distributions
relative to specific sets of two-spin quantum states. We apply the
above findings to the antiferromagnetic Heisenberg model in a uniform
magnetic field, both on a chain and on a two-leg ladder. Using Quantum
Monte Carlo simulations, we obtain the above probability distributions
and the associated entanglement, discussing their evolution under
application of the field.
}

\maketitle

\section{Introduction}
Entanglement properties have recently entered the tool kit for
studying magnetic systems, thanks to the insight they provide
on aspects which are not directly accessible through the
analysis of standard magnetic
observables
\cite{ArnesenBV01,Osterlohetal02,OsborneN02,Ghoshetal03,Vidaletal03,Verstraeteetal04,Roscildeetal04}.
The analysis of entanglement properties is particularly
indicated whenever purely quantum effects come into play, as in the
case of quantum phase transitions. However, in order to gain a deeper
insight into quantum criticality, as well as into other phenomena such
as field-induced factorization~\cite{Roscildeetal04} and saturation, the
connection between magnetic observables and entanglement estimators should
be made clearer, a goal we aim at in this paper.
On the other hand, most entanglement estimators, as defined for
quantum magnetic systems, are expressed in terms of magnetizations
and spin correlation functions. It comes therefore natural
to wonder where, inside the standard magnetic observables,
the information about
entanglement is actually stored, and how entanglement estimators
can extract it.
Quite clearly, by posing this question, one does also address the problem
of finding a possible experimental measure of entanglement,
which is of crucial relevance in developing possible solid-state devices 
for quantum computation.

In this context a privileged role is played by the concurrence $C$,
which measures the entanglement of formation between two q-bits
by an expression which is valid not only for pure
states but also for mixed ones\cite{Hilletal97,Wootters98}.
In the framework of interacting spin systems, exploiting different
symmetries of such systems the concurrence has been related to
spin-spin correlators and to
magnetizations.\cite{Wangetal02,Syljuasen03,Glaseretal03,Amicoetal04} 
However, $C$ has not yet been given a general interpretation from
the magnetic point of view, and a genuinely physical understanding
of its expression is still elusive.

Scope of this paper is therefore that of giving a simple physical 
interpretation of bipartite entanglement of formation, building a direct 
connection between entanglement estimators and occupation probabilities of
two-spin states in an interacting spin system. To this purpose we
develop a general formalism for analyzing the spin
configuration of the system, so as to directly relate it with the expression
of the concurrence. The resulting equations are then used to read
our data relative to the $S=1/2$ antiferromagnetic Heisenberg
model in a uniform magnetic field, both on a chain and on a two-leg
ladder. The model is a cornerstone in the study of magnetic
systems, extensively investigated and quite
understood in the zero-field case. When a uniform
magnetic field is applied the behavior of the system is enriched, gradually
transforming its ground state and thermodynamic behavior. 
The analysis of entanglement properties in this model,
and in particular that referring to the range of pairwise entanglement as
field increases, sheds new light not only on the physical mechanism leading 
to magnetic saturation in low-dimensional quantum systems, but also on the 
nature of some $T=0$ transitions observed in bosonic and fermionic 
systems, such as that of hard-core bosons with Coulomb interaction, and 
that described by the bond-charge extended Hubbard model, respectively.
In the former case, the connection between magnetic and bosonic model is
obtained by an exact mapping that allows a straightforward generalization
of our results to the discussion of the phase diagram of
the strongly interacting boson-Hubbard model~\cite{BruderFS93}. 
In the more complex case of the the bond-charge extended Hubbard model, 
a direct connection between the Heisenberg antiferromagnet in a
field is not formally available, but a recent work by Anfossi 
{\it et al.}\cite{Anfossietal05} has shown that some of the $T=0$ 
transitions observed in the system are characterized by long-ranged 
pairwise entanglement of the same type we observe in our magnetic model at 
saturation.

Our data result from stochastic series expansion
(SSE) quantum Monte Carlo simulations based on the directed-loop
algorithm\cite{SyljuasenS02}. The calculations
were carried on a chain with size $L=64$ and on a $L\times 2$ ladder
with $L=40$. In order to capture the ground-state behavior
we have considered inverse temperatures $\beta=2L$.

In Sec.\ref{s.magtoprob} we define the magnetic observables we refer to,
and develop the formalism which allows us to write them in terms of
probabilities for two spins to be in specific states, both at zero and at
finite temperature. In Sec.~\ref{s.spinconftoentanglement} we show how
concurrence extracts, out of the above probabilities, the specific
information on bipartite entanglement of formation. In Secs.~\ref{s.chain}
and \ref{s.ladder} we present our SSE data for the antiferromagnetic
Heisenberg model on a chain and on a square ladder respectively, and read
them in light of the discussion of Secs.~\ref{s.magtoprob} and
\ref{s.spinconftoentanglement}. Conclusions are drawn in
Sec.~\ref{s.conclusions}.

\section{From magnetic observables to spin configurations}
\label{s.magtoprob}

We study a magnetic system made of $N$ spins $S=1/2$ sitting
on a lattice. Each spin is described by a quantum operator
${\bm S}_l$, with $[S^\alpha_l,S^\beta_m]=
i\delta_{lm}\varepsilon_{\alpha\beta\gamma}S^\gamma_l$, $l$ and $m$ being
the site-indexes.

The magnetic observables we consider are the local
magnetization along the quantization axis:
\begin{equation}
M_l^z{\equiv}\left\langle S^z_l\right\rangle
\label{e.Mz}~,
\end{equation}
and the correlation functions between two spins sitting on sites
$l$ and $m$:
\begin{equation}
g_{lm}^{\alpha\alpha} \equiv \left\langle S^\alpha_l
S^\alpha_m\right\rangle \label{e.galphaalpha}~.
\end{equation}
The averages
$\left\langle~\cdot~\right\rangle$ represent
expectation values over the ground state for $T=0$, and
thermodynamic averages for $T>0$.

We now show that the above single-spin and two-spin quantities provide
a direct information on the specific quantum state of any two spins of
the system.  Let us consider the $T=0$ case first: For a lighter
notation we drop site-indexes, allowing their appearance whenever
needed.  After selecting two spins, sitting on sites $l$ and $m$, any
pure state of the system may be written as
\begin{equation}
|\Psi\rangle=\sum_{\nu\in {\cal S}}|\nu\rangle\sum_{\Gamma\in {\cal R}}
c_{\nu_\Gamma}|\Gamma\rangle~,
\label{e.Psi}
\end{equation}
where ${\cal S}$ is an orthonormal basis
for the $4$-dimensional Hilbert space of the selected spin
pair, while ${\cal R}$ is an
orthonormal basis for the
$2^{N-2}$-dimensional Hilbert space of the rest of the system.
Moreover, in order to simplify the notation, we
understand products of kets relative to (operators acting
on) different spins as tensor products,
meanwhile dropping the corresponding symbol $\otimes$.
The quantum probability for the spin pair to be
in the state $|\nu\rangle$, being the system in the pure state
$|\Psi\rangle$, is
$p_{\nu}\equiv\sum_\Gamma|c_{\nu_\Gamma}|^2$, and the
normalization condition $\langle\Psi|\Psi\rangle=1$ implies
$\sum_\nu p_{\nu}=1$.

We consider three particular bases for the spin pair:
\begin{align}
{\cal{S}}_1 {\equiv}&\{|u_{_{\rm I}}\rangle,|u_{_{\rm I\!I}}\rangle,|u_{_{\rm I\!I\!I}}\rangle,|u_{_{\rm I\!V}}\rangle\}~,
\label{e.bstandard}\\
{\cal{S}}_2 {\equiv}&\{|e_1\rangle,|e_2\rangle,|e_3\rangle,|e_4\rangle\}~,
\label{e.bBell}\\
{\cal{S}}_3 {\equiv}&\{|u_{_{\rm I}}\rangle,|u_{_{\rm I\!I}}\rangle,|e_3\rangle,|e_4\rangle\}~,
\label{e.bmixed}
\end{align}
with
\begin{align}
|u_{_{\rm I}}\rangle\equiv|\uparrow\rangle_l|\uparrow\rangle_m~&,~
  |u_{_{\rm I\!I}}\rangle\equiv|\downarrow\rangle_l|\downarrow\rangle_m~,\nonumber\\
|u_{_{\rm I\!I\!I}}\rangle\equiv|\uparrow\rangle_l|\downarrow\rangle_m~&,~
  |u_{_{\rm I\!V}}\rangle\equiv|\downarrow\rangle_l|\uparrow\rangle_m~,\nonumber\\
|e_1\rangle={\textstyle\frac{1}{\sqrt{2}}}
\left(|u_{_{\rm I}}\rangle+|u_{_{\rm I\!I}}\rangle\right)~&,~
  |e_2\rangle={\textstyle\frac{1}{\sqrt{2}}}
\left(|u_{_{\rm I}}\rangle-|u_{_{\rm I\!I}}\rangle\right)~,\nonumber\\
  |e_3\rangle={\textstyle\frac{1}{\sqrt{2}}}
\left(|u_{_{\rm I\!I\!I}}\rangle+|u_{_{\rm I\!V}}\rangle\right)~&,~
  |e_4\rangle={\textstyle\frac{1}{\sqrt{2}}}
\left(|u_{_{\rm I\!I\!I}}\rangle-|u_{_{\rm I\!V}}\rangle\right)~,\label{e.states}
\end{align}
where $|\uparrow\rangle_{l,m}(|\downarrow\rangle_{l,m})$ are eigenstates of
$S^z_{l,m}$ with
eigenvalue $+{\textstyle\frac{1}{2}}(-{\textstyle\frac{1}{2}})$.
For the coefficients entering Eq.~(\ref{e.Psi}), and for each
state $\Gamma$, the following relations hold
\begin{eqnarray}
c_{_{1\Gamma}}={\textstyle{\frac{1}{\sqrt{2}}}}\left(c_{{{_{\rm I}}}_{\Gamma}}
+c_{{{_{\rm I\!I}}}_{\Gamma}}\right)&,&
c_{_{2\Gamma}}={\textstyle{\frac{1}{\sqrt{2}}}}\left(c_{{{_{\rm I}}}_{\Gamma}}
-c_{{{_{\rm I\!I}}}_{\Gamma}}\right)~,
\label{e.c12}\\
c_{_{3\Gamma}}={\textstyle{\frac{1}{\sqrt{2}}}}\left(c_{{{_{\rm I\!I\!I}}}_{\Gamma}}
+c_{{{_{\rm I\!V}}}_{\Gamma}}\right)&,&
c_{_{4\Gamma}}={\textstyle{\frac{1}{\sqrt{2}}}}\left(c_{{{_{\rm I\!I\!I}}}_{\Gamma}}
-c_{{{_{\rm I\!V}}}_{\Gamma}}\right)~,
\label{e.c34}
\end{eqnarray}
meaning also
\begin{eqnarray}
|c_{_{1\Gamma}}|^2+|c_{_{2\Gamma}}|^2&=&
|c_{{{_{\rm I}}}_{\Gamma}}|^2+|c_{{{_{\rm I\!I}}}_{\Gamma}}|^2~,\label{e.c12+}\\
|c_{_{1\Gamma}}|^2-|c_{_{2\Gamma}}|^2&=&
|c_{{{_{\rm I}}}_{\Gamma}}c_{{{_{\rm I\!I}}}_{\Gamma}}|
\cos(\varphi^\Gamma_{{{_{\rm I}}}}-\varphi^\Gamma_{{{_{\rm I\!I}}}})~,\label{e.c12-}\\
|c_{_{3\Gamma}}|^2+|c_{_{4\Gamma}}|^2&=&
|c_{{{_{\rm I\!I\!I}}}_{\Gamma}}|^2+|c_{{{_{\rm I\!V}}}_{\Gamma}}|^2\label{e.c34+}~,\\
|c_{_{3\Gamma}}|^2-|c_{_{4\Gamma}}|^2&=&
|c_{{{_{\rm I\!I\!I}}}_{\Gamma}}c_{{{_{\rm I\!V}}}_{\Gamma}}|
\cos(\varphi^\Gamma_{{{_{\rm I\!I\!I}}}}-\varphi^\Gamma_{{{_{\rm I\!V}}}})~,
\label{e.c34-}
\end{eqnarray}
where $c_{\nu_\Gamma}\equiv|c_{\nu_\Gamma}|e^{i\varphi^\Gamma_\nu}$.

According to the usual nomenclature ${\cal{S}}_1$ and ${\cal{S}}_2$
are the {\it standard} and {\it Bell} bases, respectively, while
${\cal{S}}_3$ is here called the {\it mixed} basis. Such bases are
characterized by the fact that states corresponding to parallel and
antiparallel spins do not mix with each other. It therefore makes
sense to refer to $|u_{_{\rm I}}\rangle,|u_{_{\rm I\!I}}\rangle,|e_1\rangle$, and
$|e_2\rangle$ as {\it parallel states}, and to
$|u_{_{\rm I\!I\!I}}\rangle,|u_{_{\rm I\!V}}\rangle,|e_3\rangle$, and $|e_4\rangle$ as {\it
antiparallel states}.  The probabilities specifically related with the
elements of ${\cal{S}}_1$ will be hereafter indicated by $p_{{_{\rm I}}},
p_{{_{\rm I\!I}}}, p_{{_{\rm I\!I\!I}}}$, and $p_{{_{\rm I\!V}}}$ while $p_1,p_2,p_3,$ and $p_4$ will
be used for those relative to the elements of ${\cal{S}}_2$.  From the
normalization conditions
\begin{eqnarray}
&&p_{{_{\rm I}}}+p_{{_{\rm I\!I}}}+p_{{_{\rm I\!I\!I}}}+p_{{_{\rm I\!V}}}=1
\label{e.normstand}\\
&&p_1+p_2+p_3+p_4=1
\label{e.normBell}\\
&&p_{{_{\rm I}}}+p_{{_{\rm I\!I}}}+p_3+p_4=1~,
\label{e.normmix}
\end{eqnarray}
or equivalently from Eqs.~(\ref{e.c12+}) and (\ref{e.c34+}),
follows
$p_{{_{\rm I}}}+p_{{_{\rm I\!I}}}=p_1+p_2$, and
$p_{{_{\rm I\!I\!I}}}+p_{{_{\rm I\!V}}}=p_3+p_4$, representing
the probability for the two spins to be
parallel and antiparallel, respectively.
We do also notice that the elements of ${\cal{S}}_1$ are factorized states,
while those of ${\cal{S}}_2$ are maximally entangled ones.

The above description is easily translated
in terms of the two-site reduced density matrix
\begin{equation}
\rho=\sum_{\Gamma}
\langle\Gamma|\Psi\rangle\langle\Psi|\Gamma\rangle=
\sum_{\nu\lambda}|\nu\rangle\langle\lambda|
\sum_{\Gamma} c_{\nu_\Gamma}c_{\mu_\Gamma}^*~,
\end{equation}
whose diagonal elements are the probabilities for the
elements of the basis chosen for writing $\rho$.
The normalization conditions
Eqs.~(\ref{e.normstand}-\ref{e.normmix})
translate into ${\rm Tr}~(\rho)=1$.

Thanks to the above parametrization,
the magnetic observables~(\ref{e.Mz}) and
(\ref{e.galphaalpha}) are directly connected to the probabilities of
the two spins being in one of the states~\eqref{e.states}.
In fact it is
\onecolumn
\begin{align}
2(g^{xx}+g^{yy})=
\langle\Psi\,|S^+_lS^-_m+S^-_lS^+_m|\Psi\rangle & =\nonumber\\
&=\langle\Psi|S^+_lS^-_m+S^-_lS^+_m|\left(
|e_3\rangle\sum_\Gamma c_{_{3\Gamma}}|\Gamma\rangle+
|e_4\rangle\sum_\Gamma
c_{_{4\Gamma}}|\Gamma\rangle\right)=\nonumber\\
&=\langle\Psi|
\left(|e_3\rangle\sum_\Gamma c_{_{3\Gamma}}|\Gamma\rangle-
|e_4\rangle\sum_\Gamma c_{_{4\Gamma}}|\Gamma\rangle\right)\nonumber\\
&=(p_3-p_4)~,
\label{e.gxx+gyy.p}
\end{align}
\twocolumn
and similarly
\begin{eqnarray}
&&2(g^{xx}-g^{yy})=(p_1-p_2)~,
\label{e.gxx-gyy.p}\\
&&g^{zz}={\textstyle\frac{1}{2}}
\left(p_{{_{\rm I}}}+p_{{_{\rm I\!I}}}\right)-{\textstyle\frac{1}{4}}=
{\textstyle\frac{1}{2}}\left(p_1+p_2\right)-{\textstyle\frac{1}{4}}~,
\label{e.gzz.p}\\
&&M_z\equiv{\textstyle\frac{1}{2}}\left(M^z_l+M^z_m\right)=
\left(p_{{_{\rm I}}}-p_{{_{\rm I\!I}}}\right)~,
\label{e.Mz.p}
\end{eqnarray}
where all ${\cal{S}}_i$ are suitable to calculate $g^{zz}$, while
$(g^{xx}\pm g^{yy})$ and $M_z$ specifically require ${\cal{S}}_2$ and
${\cal{S}}_3$, respectively.
After Eqs.~(\ref{e.gxx+gyy.p})-(\ref{e.Mz.p}), one finds
\begin{eqnarray}
p_{{_{\rm I}}}&=&\textstyle{\frac{1}{4}}+g^{zz}+M_z~,
\label{e.puu}\\
p_{{_{\rm I\!I}}}&=&\textstyle{\frac{1}{4}}+g^{zz}-M_z~,
\label{e.pdd}\\
p_1&=&\textstyle{\frac{1}{4}}+g^{xx}-g^{yy}+g^{zz}~,
\label{e.p1}\\
p_2&=&\textstyle{\frac{1}{4}}-g^{xx}+g^{yy}+g^{zz}~,
\label{e.p2}\\
p_3&=&\textstyle{\frac{1}{4}}+g^{xx}+g^{yy}-g^{zz}~,
\label{e.p3}\\
p_4&=&\textstyle{\frac{1}{4}}-g^{xx}-g^{yy}-g^{zz}~.
\label{e.p4}
\end{eqnarray}
It is to be noticed that the probabilities relative to the Bell states
do not depend on the magnetization.

In the the finite temperature case, the generalization
is straightforwardly obtained by writing each of the
Hamiltonian eigenstates, numbered by the index $n$, in the form
(\ref{e.Psi}), so that
\begin{equation}
\rho(T)=
\sum_{\nu\mu}|\nu\rangle\langle\mu|
\sum_n e^{-E_n/T}\sum_{\Gamma}c_{\nu_\Gamma\!,n}c_{\mu_\Gamma\!,n}^*~.
\label{e.rho(t)}
\end{equation}
In terms of probabilities the above expression simply means that the
purely quantum $p_\mu$ shall be replaced by
the quantum statistical probabilities
\begin{equation}
p_\mu(T)\equiv\sum_n e^{-E_n/T}
\sum_{\Gamma}\left|c_{\nu_\Gamma\!,n}\right|^2~.
\label{e.p(t)}
\end{equation}
Therefore, apart from the further complication of the formalism, the
discussion developed for pure states stays substantially unchanged when
$T>0$.

Equations~(\ref{e.puu})-(\ref{e.p4}) show that magnetic observables allow a
certain insight into the spin configuration of the system, as they give,
when properly combined, the probabilities for any selected spin pair to
be in some specific quantum state. However, the mere knowledge
of such probabilities is not sufficient to appreciate
the quantum character of the global state, and more specifically
to quantify its entanglement properties.

\section{From spin configurations to entanglement properties}
\label{s.spinconftoentanglement}

We here analyze the entanglement of
formation\cite{Bennettetal96,Hilletal97,Wootters98} between two spins,
quantified by the concurrence $C$.
In the simplest case of two isolated spins in the pure state $|\phi\rangle$
the concurrence may be written as $C=|\sum_i\alpha_i^2|$, where
$\alpha_i$ are the coefficients entering the decomposition of $|\phi\rangle$
upon the magic basis $\{|e_1\rangle,i|e_2\rangle,i|e_3\rangle,|e_4\rangle\}$.
However, if one refers to the notation of the previous section,
it is easily shown that
\begin{equation}
C(|\phi\rangle) = \left|(c_{_1}^2-c_{_2}^2)-(c^2_{_3}-c^2_{_4})\right|=
2\left|c_{{_{\rm I}}}c_{{_{\rm I\!I}}}-c_{{_{\rm I\!I\!I}}}c_{{_{\rm I\!V}}}\right|~,
\label{e.Ctwospins}
\end{equation}
where Eqs.~(\ref{e.c12})-(\ref{e.c34}) have been used, with index
$\Gamma$ obviously suppressed.  The above expression shows that $C$
extracts the information about the entanglement between the two spins
by combining probabilities and phases relative to specific two-spin
state.

In fact, one should notice that a finite probability for two spins to
be in a maximally entangled state does not guarantee \emph{per se} the
existence of entanglement between them, since this probability may be
finite even if the two spins are in a separable state.\cite{esempio}
In a system with decaying correlations, at infinite separation all
probabilities associated to Bell states attain the value of $1/4$, but
this of course tells nothing about the entanglement between them,
which is clearly vanishing. It is therefore expected that
\emph{differences} between such probabilities, rather than the
probabilities themselves give insight in the presence or absence
of entanglement.


When the many-body case is tackled, the mixed-state concurrence of the
selected spin pair has an involved definition in terms of the reduced
two-spin density matrix.\cite{Wootters98} However, possible symmetries of the
Hamiltonian ${\cal{H}}$ greatly simplify the problem to the extent
that $C$ results a simple function of the probabilities
(\ref{e.puu})-(\ref{e.p4}) only. We here assume that ${\cal{H}}$ is
real, has parity symmetry (meaning that either ${\cal{H}}$ leaves the
$z$ component of the total magnetic moment unchanged, or changes it in
steps of $2$), and is further characterized by translational and
site-inversion invariance. The two latter properties implies $M^z_l$ as
defined in Eq.~(\ref{e.Mz}) to coincide with the uniform magnetization
$M_z\equiv \sum_l\langle S^z_l\rangle/N$, and the probabilities
$p_{{_{\rm I\!I\!I}}}=p_{{_{\rm I\!V}}}$, respectively.

Under these assumptions, the concurrence for a given spin pair
is\cite{Amicoetal04}
\begin{eqnarray}
C_{(r)}&{\equiv}&2\max\{0,C'_{(r)},C''_{(r)}\}~,
\label{e.Cr}\\
C'_{(r)}&{\equiv}&|g_{(r)}^{xx}+g_{(r)}^{yy}|
-
\sqrt{\left({\textstyle{\textstyle\frac{1}{4}}}+g_{(r)}^{zz}\right)^2-M_z^2}~,
\label{e.C'}\\
C''_{(r)}&{\equiv}&|g_{(r)}^{xx}-g_{(r)}^{yy}|
-{\textstyle\frac{1}{4}}+g_{(r)}^{zz}~,
\label{e.C''}
\end{eqnarray}
where $r$ is the distance in lattice units between the two selected
spins.  Despite being simple combinations of magnetic observables, the
physical content of the above expressions is not straightforward.
However, by using the expression found in Section \ref{s.magtoprob},
one can write Eqs.~\eqref{e.C'} and \eqref{e.C''} in terms of the
probabilities for the two spins to be in maximally entangled or
factorized states, thus finding, in some sense, an expression which is
analogous to Eq.~\eqref{e.Ctwospins} for the case of mixed
states. In fact, from Eqs.~(\ref{e.gxx+gyy.p})-(\ref{e.gxx-gyy.p}), it
follows
\begin{eqnarray}
2C'&=&|p_3-p_4|-2\sqrt{p_{{_{\rm I}}} p_{{_{\rm I\!I}}}}\label{e.C'p}~,\\
2C''&=&|p_1-p_2|-(1-p_1-p_2)=\nonumber\\
&=&|p_1-p_2|-2\sqrt{p_{{_{\rm I\!I\!I}}} p_{{_{\rm I\!V}}}}\label{e.C''p}~,
\end{eqnarray}
where we have used $p_{{_{\rm I\!I\!I}}}{=}p_{{_{\rm I\!V}}}$ and hence
$p_3+p_4=2p_{{_{\rm I\!I\!I}}}=2\sqrt{p_{{_{\rm I\!I\!I}}}p_{{_{\rm I\!V}}}}$.
The expression for $C''$ may be written in the particularly simple form
\begin{equation}
2C''=2{\rm max}\{p_1,p_2\}-1~,
\label{e.C''pmax}
\end{equation}
telling us that, in order for $C''$ to be positive,
it must be either $p_1>1/2$ or $p_2>1/2$. This means that
one of the two parallel Bell states needs to saturate at least half
of the probability, which implies that it is by far the state
where the spin pair is most likely to be found.

Despite the apparently similar structure of Eqs.~(\ref{e.C'p}) and
(\ref{e.C''p}), understanding $C'$ is more involved, due to the fact that
$\sqrt{p_{{_{\rm I}}}p_{{_{\rm I\!I}}}}$ cannot be further simplified unless
$p_{{_{\rm I}}}=p_{{_{\rm I\!I}}}$. The marked difference between
$C'$ and $C''$ reflects the different mechanism through which
parallel and antiparallel entanglement is generated when time reversal
symmetry is broken, meaning $p_{{_{\rm I}}}\neq p_{{_{\rm I\!I}}}$ and hence $M_z\neq 0$.
In fact, in the zero magnetization case, it is
$p_{{_{\rm I\!I}}}=p_{{_{\rm I}}}=(p_1+p_2)/2$ and hence
\begin{equation}
2 C'=2{\rm max}\{p_3,p_4\}-1~,
\label{e.C'p.Mz=0}
\end{equation}
which is fully analogous to Eq.~(\ref{e.C''pmax}), so that the above
analysis can be repeated by simply replacing $p_1$ and $p_2$ with $p_3$ and
$p_4$.

For $M_z \neq 0$, the structure of Eq.~(\ref{e.C'p.Mz=0}) is somehow kept by
introducing the quantity
\begin{equation}
\Delta^2\equiv(\sqrt{p_{{_{\rm I}}}}-\sqrt{p_{{_{\rm I\!I}}}})^2~,
\label{e.Delta2}
\end{equation}
so that
\begin{equation}
2 C'=2{\rm max}\{p_3,p_4\}-(1-\Delta^2)~,
\label{e.C'Delta}
\end{equation}
meaning that the presence of a magnetic field favors bipartite
entanglement associated to antiparallel Bell states, $|e_3\rangle$ and
$|e_4\rangle$. In fact, when time reversal symmetry is broken the
concurrence can be finite even if $p_3,~p_4<1/2$.

From Eqs.~(\ref{e.C''pmax}) and (\ref{e.C'Delta}) one can conclude that,
depending on $C$ being finite due to $C'$ or $C''$, the entanglement
of formation originates from finite probabilities for the two selected
spins to be parallel or antiparallel, respectively. In this sense we
will speak about {\it parallel} and {\it antiparallel}
entanglement.

Moreover, from Eqs.~(\ref{e.C'p})-(\ref{e.C''p}) we notice that, in
order for parallel (antiparallel) entanglement to be present in the
system, the probabilities for the two parallel (antiparallel) Bell
states must be not only finite but also different from each
other. Thus, the Bell states $|e_1\rangle$ and $|e_2\rangle$
($|e_3\rangle$ and $|e_4\rangle$) result mutually exclusive in the
formation of entanglement between two spins in the system, the latter
being present only if one of the Bell state is more probable than the
others. The case $p_1=p_2=1/2$ ($p_3=p_4=1/2$) corresponds in turn to
an incoherent mixture of $|e_1\rangle$ and $|e_2\rangle$
($|e_3\rangle$ and $|e_4\rangle$).

In fact, the occurrence of the differences $|p_1-p_2|$ and $|p_3-p_4|$
is intriguing. Let us comment on $|p_1-p_2|$, as the same kind of
analysis holds for $|p_3-p_4|$.  In the general case the difference
$p_1-p_2$ can vanish because of genuine many-body effects which are
not directly readable in terms of 2-spin entangled or separable states.
It is easier to interpret Eq.~\eqref{e.C''p} [Eq.~\eqref{e.C'p}], if one
restricts the possibilities to the case in which the two spins are not
entangled with the rest of the system. By
using Eq.~(\ref{e.c12-}), one can select two particular situations all
leading to $p_1=p_2$:\\
\noindent
{\it (i)} $c_{{{_{\rm I}}}_{\Gamma}}$ or $c_{{{_{\rm I\!I}}}_{\Gamma}}$
vanishes $\forall\Gamma$, meaning that $|\Psi\rangle$ does
not contain states where the two selected spins are parallel and
entangled;\\
\noindent
{\it (ii)} for each $\Gamma$ such that both $|c_{{{_{\rm I}}}_{\Gamma}}|$ and
$|c_{{{_{\rm I\!I}}}_{\Gamma}}|$ are non-zero, it is $\varphi_{_{\rm
I}}^\Gamma-\varphi_{_{\rm II}}^\Gamma=\pi/2$. Thus, whatever the
antiparallel components are, the parallel terms of $|\Psi\rangle$
appear in the form $(\alpha |e_1\rangle+\alpha^*|e_2\rangle)$.

The above analysis suggests the first term in $C''$ ($C'$)
to distill, out of all possible parallel (antiparallel) spin
configurations, those which are specifically related with entangled
parallel (antiparallel) states. These characteristics reinforce the
meaning of what we have called parallel and antiparallel entanglement.

\section{Chain}
\label{s.chain}

We consider the isotropic Heisenberg antiferromagnetic chain in a uniform
magnetic field, described by
\begin{equation}
\frac{\cal{H}}{J}=
\sum_i
\boldsymbol{S}_i\cdot\boldsymbol{S}_{i+1}-h S^z_i~,
\label{e.chain}
\end{equation}
where the exchange integral $J$ is positive, and the
reduced magnetic field $h{\equiv}g\mu_{\rm B}H/J$ is assumed
uniform.

This model is characterized by the rotational symmetry on the $xy$
plane, as well as by the existence of a saturation field
$h_{\rm s}=2$, such that for $h\geq h_{\rm s}$ the ground state is the
factorized ferromagnetic one, with all spins aligned along the field
direction.
Moreover, Eq.~(\ref{e.chain}) has all the necessary symmetries for
Eqs.~(\ref{e.Cr})-(\ref{e.C''}) to hold.

Due to the rotational symmetry on the $xy$ plane, it is
$g^{xx}=g^{yy}$, meaning $p_1=p_2=\frac14 + g^{zz} \leq 1/2$, according to
Eqs.~(\ref{e.p1}) and (\ref{e.p2}), and hence null parallel entanglement
($C'' \leq 0$) between any two spins along the chain, no matter
the field, the temperature, and the distance between them.

\begin{figure}
\resizebox{0.75\columnwidth}{!}{\includegraphics{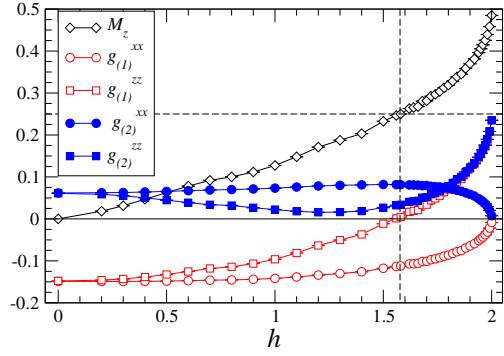}}
\caption{Magnetization and correlators versus the
magnetic field $h$ for the chain Eq.~\eqref{e.chain}. The dashed lines mark 
the value of the field where $p_{{_{\rm I\!I}}}=0$ (see text).}
\label{f.corr.chain} 
\end{figure}
In Fig.~\ref{f.corr.chain} we show the $T=0$ correlation functions for
nearest neighboring (n.n.) and next-nearest neighboring (n.n.n.)
spins, together with the uniform magnetization, as the field is
varied. Beyond the overall regular behavior, we notice that there
exists a value of the magnetic field where one simultaneously observes
$g_{(1)}^{zz}=0$ and $M_z=1/4$ (indicated by the dashed
lines). According to Eqs.~(\ref{e.gzz.p}) and (\ref{e.Mz.p}) this
implies null probability $p_{{{}_{\rm I\!I}}}$ for adjacent spins to
be parallel in the direction opposite to the field. This means that
the ground-state configuration is a superposition of spin
configurations entirely made of stable clusters of spins parallel to
the field separated by N\'eel-like strings.

\begin{figure}
\resizebox{0.75\columnwidth}{!}{\includegraphics{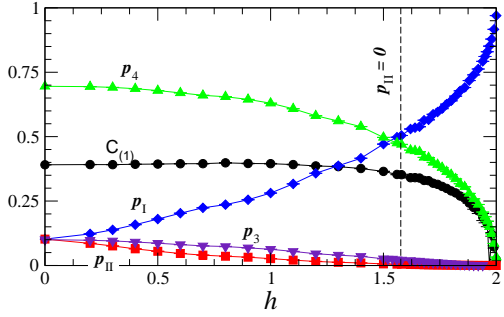}}
\caption{Concurrence and related probabilities of the mixed basis states ${\cal
S}_3$ [Eq.~\eqref{e.bmixed}] for n.n. sites of the chain
Eq.~\eqref{e.chain}}
\label{f.C1prob.chain}
\end{figure}

 In Fig.~\ref{f.C1prob.chain} we show the probabilities for
n.n. spins to be in the states of the mixed basis, together with
the n.n. concurrence: 
The value of the n.n. concurrence for $h=0$ is in agreement with the
exact resut in the thermodynamic limit.\cite{Glaseretal03}
In presence of an external magnetic field, 
$C_{(1)}$ is found positive $\forall h$, meaning
that, no matter the value of the field, the probabilities $p_3$ and
$p_4$ for adjacent spins are always different from
each other. The probabilities for the triplet states
$|e_3\rangle,|u_{{_{\rm I\!I}}}\rangle,$ and $|u_{{_{\rm I}}}\rangle$ are
equal for $h=0$ and depart from each other when the field is switched
on. The singlet $|e_4\rangle$ evidently dominates the ground state
up to a field which roughly corresponds to the value where
$p_{{_{\rm I\!I}}}$ vanishes.

As for the concurrence, despite the ground-state structure
evidently changes as the field increases, $C_{(1)}$ stays substantially
constant up to
a large value of the field, mainly due to the fact that not only $p_4$
but also $p_3$ decreases with the field. This behavior mimics the one
occurring in a spin dimer, whose ground state is the singlet state
$|e_4\rangle$ up to $h=1$ where, after a level crossing,
$|u_{{_{\rm I}}}\rangle$ becomes energetically favored.
However, in a spin chain, many-body effects smear the sharp behavior of the
dimer due to the level
crossing.  We do also notice that $C_{(1)}$ starts to
decrease as soon as the total probability for parallel spins
($p_{{_{\rm I}}}+p_{{_{\rm I\!I}}}$) gets larger than that for antiparallel spins
($p_3+p_4$). The further reduction of $C_{(1)}$ is mainly driven by
$p_{{_{\rm I}}}$ starting to rapidly increase.

\begin{figure}
\resizebox{0.75\columnwidth}{!}{\includegraphics{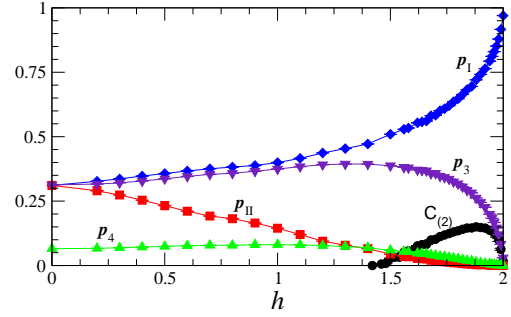}}
\caption{Concurrence and related probabilities of the mixed basis states ${\cal
S}_3$ [Eq.~\eqref{e.bmixed}] for n.n.n. sites of the chain
Eq.~\eqref{e.chain}.}
\label{f.C2prob.chain}
\end{figure}

In the same field region where a substantial change in the
n.n. configuration occurs, the n.n.n. concurrence $C_{(2)}$ switches
on.  This is seen in Fig.~\ref{f.C2prob.chain}, where the
probabilities for n.n.n. spins are shown together with the
corresponding concurrence. In fact, when considering the
n.n.n. quantities, we notice that both $g_{(2)}^{xx}$ and
$g_{(2)}^{zz}$ have a non-monotonic behavior, displaying a maximum and
a minimum, respectively, in the field region where $C_{(2)}$ gets
positive (as from the comparison between Fig.~\ref{f.corr.chain} and
\ref{f.C2prob.chain}).

Regarding the probabilities, one finds that, although the most likely
state is always $|u_{{_{\rm I}}}\rangle$, $p_3$ is surprisingly large, and
almost equal to $p_{{_{\rm I}}}$, as far as $h<1$. Moreover, both $p_3$ and
$p_4$ have a non monotonic behavior and increase with $h$ up to the
field where we simultaneously observe $g_{(2)}^{xx}$ and
$g_{(2)}^{yy}$ attaining their extreme values, $p_{{_{\rm I}}}$ exceeding
$1/2$, $p_4$ getting larger than $p_{{_{\rm I\!I}}}$, and $C_{(2)}$ switching
on.

As observed in the n.n. case, when $p_{{_{\rm I\!I}}}$ for n.n.n. spins vanishes
$C_{(3)}$ switches on.  Let us further comment upon
$C_{(1)}$, $C_{(2)}$, and $C_{(3)}$. Given the fact that only
antiparallel entanglement may exist in this chain, it is not
surprising that $C_{(1)}>0$ and $C_{(2)}=0$ at low fields, as
n.n. spins belong to different sublattices, while n.n.n. spins belong
to the same sublattice. However, the fact that $C_{(2)}$ becomes
finite indicates a ground-state evolution from the N\'eel-like to the
ferromagnetic state such that the system enters a region where quantum
fluctuations increase the total probability for spins belonging to the
same sublattice to be antiparallel and entangled.  The opposite effect
is understood when $C_{(3)}$ is considered: in order to keep
$C_{(3)}=0$ almost up to the saturation field, quantum fluctuations
must reduce the total probability for spins belonging to different
sublattices to be antiparallel and entangled.

\begin{figure}
\resizebox{0.75\columnwidth}{!}{\includegraphics{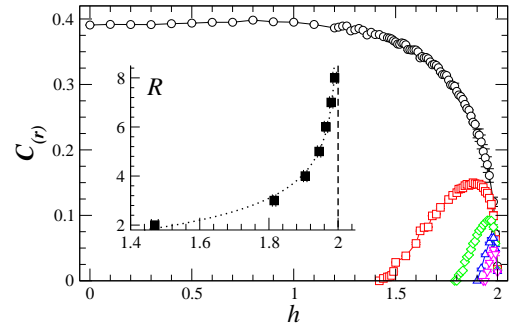}}
\caption{From the upper to the lower curves:
concurrences from the nearest- up to the $5^{\rm th}$-neighbors of the
chain Eq.~\eqref{e.chain}. The inset shows the divergence of the range of
the concurrence as $h\to h_{\rm s}$, the line shows the $(h-h_{\rm
s})^{-1/2}$ behavior.}
\label{f.Cn.chain} 
\end{figure}
The above comments upon $C_{(2)}$ and $C_{(3)}$ may be generalized to
$C_{(n)}$ with even and odd $n$, respectively. In
Fig.~\ref{f.Cn.chain} we in fact show $C_{n}$ up to $n=5$. The
concurrence for increasing distance between the two spins gets finite
for a big enough field resembling the phenomenology of finite spin
clusters.\cite{ArnesenBV01} Moreover, combining the exact results of
Refs.\cite{Jinetal04} and \cite{Hikiharaetal04}, we find
that the range of the concurrence for the model~\eqref{e.chain},
namely the distance $R$ such that
$C_{(r)}$ vanishes for $r>R$, is
\begin{equation}
R=\left|\frac{\rho}{\sqrt{\pi}(2+4M_z)(\frac12 - M_z)^{1/2}}
\right|^\theta~,
\label{e.Crange}
\end{equation}
with the constant $\rho=0.924...$.  When $h\to h_{\rm s}$, it is $M_z
\simeq \frac12 - \frac{\sqrt{2}}{\pi}\sqrt{h_{{\rm s}}-h}$ and $\theta
\simeq 2 - \frac{2\sqrt{2}}{\pi}\sqrt{h_{{\rm s}}-h}$, and the range
of the concurrence is seen to diverge according to $R\simeq
\frac{\rho\sqrt2}{32}(h_{{\rm s}}-h)^{-1/2}$.
In other terms, approaching the saturation field, all $C_{(n)}$ become
finite of order $O(1/N)$, consistently with the occurrence of a
$|W_N\rangle$ state\cite{Duretal00}.  For such state the entanglement
is maximally bipartite in the sense of the Coffman-Kundu-Wootters
inequality.\cite{Duretal00,Coffmanetal00,Osborne05} This scenario is
consistent with our numerical data. As shown in Fig.\ref{f.Cn.chain}
up to $n=5$, for any $C_{(n)}$ it exists a field $h_n>h_{n-1}$ such
that $C_{(n)}$ is positive for $h\in[h_n,2)$, with $h_n\to 2$ for
$n\to\infty$. The divergence of the range of the concurrence for $h\to
h_{{\rm s}}$ is shown in the inset of Fig.~\ref{f.Cn.chain}. Although
the correct power-law behavior shows up, the precision of the
numerical data is not sufficient to get the correct multiplicative
constant. In fact, the above expression~\eqref{e.Crange} is derived
from asymptotic exact results, valid only for $r\gg 1$, when $C_{(r)}$
becomes too small to resolve it numerically.

\begin{figure}
\resizebox{0.80\columnwidth}{!}{\includegraphics{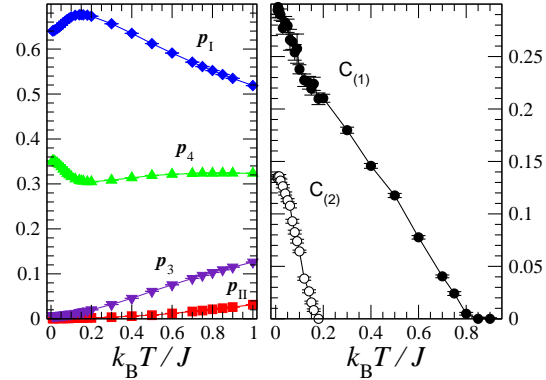}}
\caption{Left panel: n.n. probabilities versus temperature at
$h=1.8$. Right panel: n.n. and n.n.n. concurrences versus temperature
at $h=1.8$.}
\label{f.temp}
\end{figure}

The formalism introduced in the previous sections works also in the
finite temperature case, where it describes the effects of thermal
fluctuations on quantum coherence. In Fig.~\ref{f.temp}, the
temperature dependence of probabilities and concurrences, for $h=1.8$,
shows how thermal fluctuations progressively drive the system towards
an incoherent mixture of states. Increasing $T$ the concurrences
(right panel) are progressively suppressed and above $k_{_{\rm
B}}T\sim0.8J$ also the n.n. concurrence vanishes. At higher
temperatures none of the spin pairs in the system is entangled and
quantum coherence is lost. The temperature behavior of the
probabilities (left panel) is non monotonic, signaling the relative
weight of the different states in the energy spectrum of the
system. Eventually, at high $T$ all the probabilities tends to the
asymptotic value $p_\nu=1/4$.

\section{Two-Leg Ladder}
\label{s.ladder}

The above picture further enriches when considering the two-leg
isotropic ladder, described by
\begin{equation}
\frac{\cal{H}}{J}=\sum_i\sum_{\alpha=0,1}
(\boldsymbol{S}_{i,\alpha}\cdot\boldsymbol{S}_{i+1,\alpha}{-}
hS^z_{i,\alpha})
+\gamma\boldsymbol{S}_{i,{0}}\cdot\boldsymbol{S}_{i,{1}}~,
\label{e.hHeisenberg.ladder}
\end{equation}
where the index $i$ runs on both the right ($\alpha=0$) and left
($\alpha=1$) leg. The first term is the Heisenberg
Hamiltonian\,\eqref{e.chain} for the right and left legs, while the last
term describes the exchange interaction between spins of the same
rung, whose relative weight is $\gamma$.

The model\,\eqref{e.hHeisenberg.ladder} is known to describe cuprate
compounds like SrCu$_2$O$_3$ and it has been extensively studied for
zero\cite{Barnesetal93,Dagottoetal96} and finite
field\cite{Chaboussantetal98}.  The system shows a gap $\Delta$ in the
excitation spectrum that can be interpreted essentially as due to
the energy cost for producing a triplet excitation on a rung
\cite{Barnesetal93}. The system reaches full
polarization\cite{Chaboussantetal98}, with all spins aligned along the
field direction, for $h > \gamma + 2$.

In the following we will specifically consider the isotropic case $\gamma=1$,
which is characterized by a gap $\Delta \approx 0.5 J$,\cite{Barnesetal93}
and by a saturation field $h_{\rm s}=3$.
As in the chain case, due to the rotational invariance on the $xy$ plane,
parallel entanglement cannot develop in the isotropic ladder. On the other
hand, antiparallel bipartite entanglement can here develop between
spins belonging to the same leg, or to the same rung, or to a
different rung and leg. Two-spin quantities will be hereafter pinpointed by
the two-component vector $(r_i,r_\alpha)$ joining the two selected spins,
the first component
referring to the direction of the legs, and the second one to that of the
rungs. The indexes $(01)$, $(10)$, $(11)$, $(20)$
will therefore indicate n.n. spins on the same rung, n.n. along one leg,
n.n.n. on adjacent rungs, and n.n.n.
along the same leg, respectively.

\begin{figure}
\resizebox{0.75\columnwidth}{!}{\includegraphics{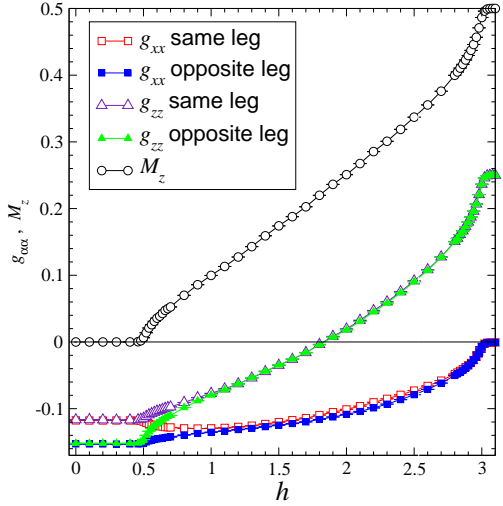}}
\caption{Magnetization and correlators versus the magnetic field
$h$ for the ladder Eq.~\eqref{e.hHeisenberg.ladder}.}
\label{f.corr.ladder}
\end{figure}
Our SSE data in Fig.~\ref{f.corr.ladder} for the uniform
magnetization and the n.n. correlation functions
$g^{\alpha\alpha}_{(01)}$, $g^{\alpha\alpha}_{(10)}$ confirm the
description given in the previous paragraph:
Before the Zeeman interaction fills the energy gap
at the critical value $h_{\rm c}\simeq 0.5$, the ground-state configuration
is frozen and characterized by the singlet $|e_4\rangle$ being by far
the most likely state for each rung.

\begin{figure}
\resizebox{0.75\columnwidth}{!}{\includegraphics{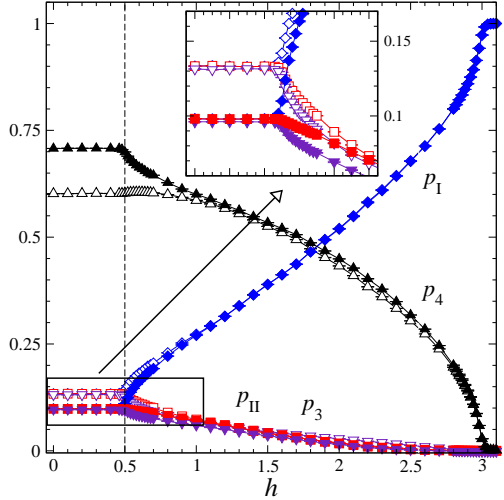}}
\caption{Probabilities relative to the mixed basis versus the magnetic field
$h$ for the ladder Eq.~\eqref{e.hHeisenberg.ladder}: $p_{{_{\rm I}}}$
($\Diamond$), $p_{_{\rm I\!I}}$ ($\Box$), $p_3$ ($\triangle$), and
$p_4$ ($\nabla$). Open (full) symbols are for n.n. spins along the
same leg (on the same rung). The inset zooms in on the behavior of
$p_{_{\rm I}}$, $p_{_{\rm I\!I}}$, and $p_3$ near the critical field
$h_{\rm c}$.}
\label{f.prob.ladder}
\end{figure}
The use of the formalism developed in Sec.~\ref{s.magtoprob} gives a
direct information on the physics of the system: In
Fig.~\ref{f.prob.ladder} we see that the singlet probabilities $p_4$
relative to n.n. spins on a rung and along one leg, as functions of the
field, share a similar behavior everywhere but at the critical field
$h_c$, where $p_4$ for n.n. spins sitting on the same rung shows up a kink
that is not present in the singlet probability along the leg. This
qualitatively different behavior clearly reflects the nature of the energy
gap that closes at $h_c$. The sharp decrease of $p_4$ in favor of
$p_{{_{\rm I}}}$ on the rung just above $h_c$ testifies that, even in the case,
here considered, of equal exchange interaction along the legs and on the
rungs ($\gamma=1$), the first excitations in the energy spectrum of the
ladder are triplet excitations on the rungs.

When the Zeeman energy becomes larger than the gap, for $h>h_{\rm c}$,
the ground state starts to evolve with the field, whose immediate
effect is that of pushing the quantities relative to spins on the
rungs and along the legs towards each other: In fact, for $h>1$
n.n. spins along the legs and on the rungs substantially share the
same behavior. As for the probabilities, we see that $p_{{_{\rm
I\!I}}}$ and $p_3$ keep being equivalent, no matter the value of the
field, and slowly vanish as saturation is reached.  On the contrary
the probability for n.n. spins to be in $|u_{{_{\rm I}}}\rangle$
increases at the expense of the probability relative to the singlet
state until, for $h\simeq 1.8$, the two probabilities cross each
other.

\begin{figure}
\resizebox{0.75\columnwidth}{!}{\includegraphics{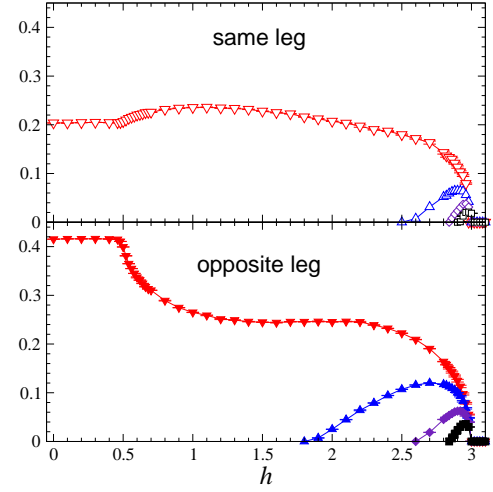}}
\caption{From the upper to the lower curves:
concurrences relative to spins belonging to the same (upper panel) and
different legs (lower panel) versus field for the isotropic ladder up to
$r=4$.}
\label{f.Cn.ladder} 
\end{figure}

Finally, we apply the formalism of
Sec.~\ref{s.spinconftoentanglement} to extract features of the ground
state from the concurrences. Fig.~\ref{f.Cn.ladder} shows the
concurrences $C_{(r_i,r_\alpha)}$ up to the distance
$r\equiv r_i+r_\alpha=4$ for spins sitting on the same (upper panel)
and on different legs. The bipartite antiparallel entanglement between
two spins sitting at a given distance $r$ is in general larger on
different legs, even beyond the n.n. case.

As expected, the field, after closing the gap, pushes $C_{(01)}$ and
$C_{(10)}$ towards each other. Quite unexpectedly, however, this
evolution includes a region where n.n. concurrence along the leg,
$C_{(10)}$, increases. It is interesting to notice that
$C_{(11)}$ switches on at $h\simeq 1.8$, where $p_4$ and $p_{{_{\rm I}}}$ for
n.n. spins are seen to cross each other in Fig.~\ref{f.corr.ladder},
signaling the crossover from an antiferromagnetic to a
ferromagnetic-like configuration of the n.n. spins.

\section{Conclusions}
\label{s.conclusions}

In this paper we developed a simple and effective formalism that
allows to reconstruct the probability for two spins of a multi-spin
system to be in a given quantum state, once the collective state of
the system is given. Remarkably, such probabilities are found to be
simple combination of standard magnetic observables,
Eqs.~(\ref{e.puu}-\ref{e.p4}). Within such formalism it is very
natural to understand how concurrence quantifies the amount of
entanglement between two spins by comparing the probabilities for
those spins to be in different Bell states. In particular the
expression for the concurrence clearly separates the case of parallel
[Eq.~\eqref{e.C''pmax}] and antiparallel [Eq.~\eqref{e.C'Delta}]
spins, leading to the introduction of the concept of \emph{parallel}
and \emph{antiparallel} entanglement.

The knowledge of the probability distribution for a given set of two-spin
states can be a useful tool to study quantum phases
dominated by the formation of particular local two-spin states
and to investigate the transitions given by the alternation
of such states. Within this class of phenomena we can cite
the occurrence of short-range valence-bond states in low-dimensional
quantum antiferromagnets \cite{Whiteetal94}, and the transition
from a dimer-singlet phase to long-range order in systems
of weakly coupled dimers under application of a field
or by tuning of the inter-dimer coupling \cite{Matsumotoetal04}.

\section{Acknowledgments}

 Fruitful discussions with L. Amico, G.~Falci, S. Haas, D.~Pa\-ta\-n\`e,
 J.~Siewert, and R.~Vaia are gratefully acknowledged. We acknowledge
 support by SQUBIT2 project EU-IST-2001-390083 (A.F.), and by NSF
 under grant DMR-0089882 (T.R.).


\begin{thebibliography}{200}
\bibitem{esempio}
For instance, the factorized state $|u_{{_{\rm I}}}\rangle$ can be written as
$|u_{{_{\rm I}}}\rangle=(|e_1\rangle+|e_2\rangle)/\sqrt{2}$, implying
$p_1=p_2=1/2$.
\end{thebibliography}
\end{document}